\documentclass[11pt,onecolumn,a4paper]{quantumarticle}

\newif\ifARXIV
\newif\ifUSEBBL

\ARXIVtrue

\USEBBLtrue

\pdfoutput=1
\usepackage[font={small}]{caption}
\usepackage{subfigure}
\usepackage{graphicx}
\usepackage[numbers,sort&compress]{natbib}
\usepackage{float}
\usepackage[normalem]{ulem}
\usepackage{mathrsfs} 
\usepackage{multicol}
\usepackage{cancel}
\usepackage{mathtools}
\usepackage{hyperref}
\usepackage{xcolor}
\usepackage{simplewick}
\usepackage{amsmath}
\usepackage{amsfonts}
\usepackage{amssymb}
\usepackage{graphicx}
\usepackage{simplewick}
\usepackage{hyperref}
\usepackage{multirow}
\usepackage{enumitem}
\usepackage{physics}
\usepackage{booktabs}
\usepackage{xcolor}
\usepackage{dsfont}
\usepackage{subfigure}
\usepackage{amsmath,amsthm,appendix,bm,graphicx,hyperref,mathrsfs,setspace,slashed,xcolor}
\usepackage{quantikz,stmaryrd}
\def\Z2{ {\mathbb Z}_{2} }

\newcommand{\KET}[1] {| #1 \rangle}

\def\ALLNQI{PhysRevLett.81.3992,150706334,BechmannPLA98,9802051,9803019,9811036,0309189,0502072,BrunPRL09,BennettPRL09,13033537,13030371,13107301,200907800,220613362,XuPRR22,211105977,240416288,240310102,250306423}
\def\VV{Valiant-Vazirani }

\begin{document}


\title[]{Quantum algorithm for Valiant-Vazirani reduction}

\author{Patrick Kelly}
\affiliation{School of Computing, University of Georgia, Athens, Georgia 30602, USA}
\affiliation{Center for Simulational Physics, University of Georgia, Athens, Georgia 30602, USA}

\author{Victoria S. Ordonez}
\affiliation{School of Computing, University of Georgia, Athens, Georgia 30602, USA}
\affiliation{Center for Simulational Physics, University of Georgia, Athens, Georgia 30602, USA}

\author{Michael R. Geller}
\affiliation{Department of Physics and Astronomy, University of Georgia, Athens, Georgia 30602, USA}
\affiliation{Center for Simulational Physics, University of Georgia, Athens, Georgia 30602, USA}

\author{Yohannes Abate}
\affiliation{Department of Physics and Astronomy, University of Georgia, Athens, Georgia 30602, USA}

\date{June 15, 2026}

\begin{abstract}
There is growing interest in extensions of the standard model of gate-based quantum computation to include auxiliary degrees of freedom evolving according to a nonlinear Schr\"odinger equation.  By reducing the Boolean satisfiability problem SAT to quantum state discrimination, Abrams and Lloyd argued that the right type of nonlinearity can be used to solve {\sf NP} and \#{\sf P} problems in polynomial time, at least in an idealized noise-free limit. For practical implementation, however, we are restricted to simulated and emergent nonlinearities, such as that appearing in mean field models for ultracold atoms and similar ensembles. A prominent example is the torsion model, which arises in two-component Bose-Einstein condensates and spin models with all-to-all Ising interaction. But torsion-based state discrimination appears to fall short  of solving SAT. Here we close this gap by constructing the filtered oracle of the Valiant-Vazirani theorem, providing a randomized polynomial-time reduction from SAT to UNIQUE SAT, a promise problem where there is  at most 1 satisfying assignment. In the noise free limit, the UNIQUE SAT  problem can be  solved in polynomial time using torsion nonlinearity. Quantum Valiant-Vazirani reduction is no faster than the efficient classical version,  but a fault-tolerant implementation coupled to a nonlinear quantum coprocessor simulating torsion would enable polynomial time solution to {\sf NP} (but not \#{\sf P}) problems.
\end{abstract}

\maketitle
\setcounter{tocdepth}{2}
\clearpage


\section{Nonlinear quantum computation}

A growing body of theoretical work has 
explored the use of nonlinearity to enhance 
quantum information processing \cite{\ALLNQI}.
This is distinct from the challenge of designing (linear) quantum algorithms to speed up the simulation of nonlinear phenomena \cite{GaitanNPJQI20,201103185,201106571,211212307,240606323}. Abrams and Lloyd \cite{PhysRevLett.81.3992} gave an efficient reduction (in the form of a quantum circuit) from any problem in the complexity class {\sf NP} to the problem of discriminating exponentially close states of a single qubit.  If we restrict ourselves to the standard circuit model, the Abrams-Lloyd reduction achieves no speedup (because the required state discrimination would not be efficient). However if certain types of nonlinear operations are allowed on that one qubit, exponential speedup for the Boolean satisfiability problem  SAT is possible in the noise-free limit.  If implemented on a scalable and fault-tolerant platform, this would allow nonlinear quantum computers to solve any {\sf NP} problem in polynomial time, which is believed to be beyond the reach of {\sf BQP} quantum computation \cite{BennettBernstein9701001}.
With additional assumptions, Abrams and Lloyd
proposed that \#{\sf P} problems could be solved in  polynomial time as well \cite{PhysRevLett.81.3992}. Beyond single-qubit nonlinearity, Aaronson has argued that a noise-free quantum computer with certain nonlinear AND and OR gates can solve any {\sf PSPACE} problem in polynomial time \cite{0502072}.
 
The theoretical evidence for nonlinear speedup seems strong, but how do we use it?
One approach is through analog quantum simulation: Design a controllable quantum system 
to simulate a target nonlinear model.
However the use of simulated nonlinearity raises two issues:  (i) How can speedup be obtained if the system is truly linear?  The reason is that an accurate simulation requires large $N$, where $N$ is the number of particles or spins. The simulator does achieve exponential time speedup at the expense of exponential space cost. Since $N$ cannot be made exponentially large in practice, there will be $O(1/N)$ errors that have to be corrected 
along with gate and decoherence errors.
(ii) The second question is how to make such a simulator. Most of the 
previous work has focused on the simulation of a single qubit with torsion nonlinearity \cite{MielnikJMP80,150706334,211105977,240416288,241022032}:
\begin{equation}
H = g \langle \sigma^z \rangle  \sigma^z , \ \ 
\langle \sigma^z \rangle  := {\rm tr}(\rho \sigma^z), \ \  g \in {\mathbb R}.
\label{torsion hamiltonian}
\end{equation}
Here $\rho$ is the current state. 
This Hamiltonian generates a $z$ rotation on the Bloch sphere with frequency $2g z$ proportional to the $z$ coordinate, twisting the Bloch sphere about the $z$ axis.  Torsion arises naturally in two-mode Bose-Einstein condensates described by the two-component Gross-Pitaevskii equation \cite{150706334}. 
Spinor and toroidal Bose-Einstein condensate 
 simulators have been proposed \cite{241022032,240416288}. 
The torsion model also arises in the large $N$ limit of the Kitagawa-Ueda model \cite{KitagawaUedaPRA1993}. Recently, 
Luo {\it et al.}~\cite{240219429} 
simulated  the model 
(\ref{torsion hamiltonian})
as well as more complex nonlinear models in an optical cavity platform.

Torsion-based nonlinear state discrimination has been investigated by Childs and Young \cite{150706334} and by Geller \cite{241022032}. 
Both papers give designs for fast single-input state discrimination of exponentially close qubit states, an impossibility within linear one-particle quantum mechanics, but fall short of achieving a 
polynomial-time solution to an {\sf NP}-complete problem. This is due in part to the exponentially large range of $s $, the number of satisfying assignments in SAT on $n$ bits, ranging from 0 to $2^n$.

In this paper we close this gap by  giving a 
quantum implementation of 
\VV reduction.
The \VV  theorem \cite{ValiantVaziraniTCS1986} is a landmark result in computational complexity. It provides a randomized polynomial-time reduction from the standard SAT problem to UNIQUE SAT, a promise problem with at most 1 satisfying assignment \cite{BlassGurevichIC1982}. The theorem shows that SAT, which is {\sf NP}-complete, remains 
{\sf NP}-hard under randomized reductions even if the input instances are promised to have at most one satisfying assignment. In the noise free limit,  the UNIQUE SAT  problem can be solved in polynomial time using torsion \cite{150706334,241022032,260514822}. 
For UNIQUE SAT, the Abrams--Lloyd encoding reduces the problem to distinguishing the cases $s=0$ and $s=1$ on a single ancilla qubit. The discriminator first applies a $y$ rotation to place the Bloch vectors representing $s=0$ and $s=1$ on opposite sides of the equator, then evolves them under the torsion Hamiltonian $H=B\sigma^x+g\langle\sigma^z\rangle\sigma^z$, driving them to opposite poles in polynomial time \cite{150706334,241022032,260514822}.

\begin{figure}
\begin{center}
\includegraphics[width=14.0cm]{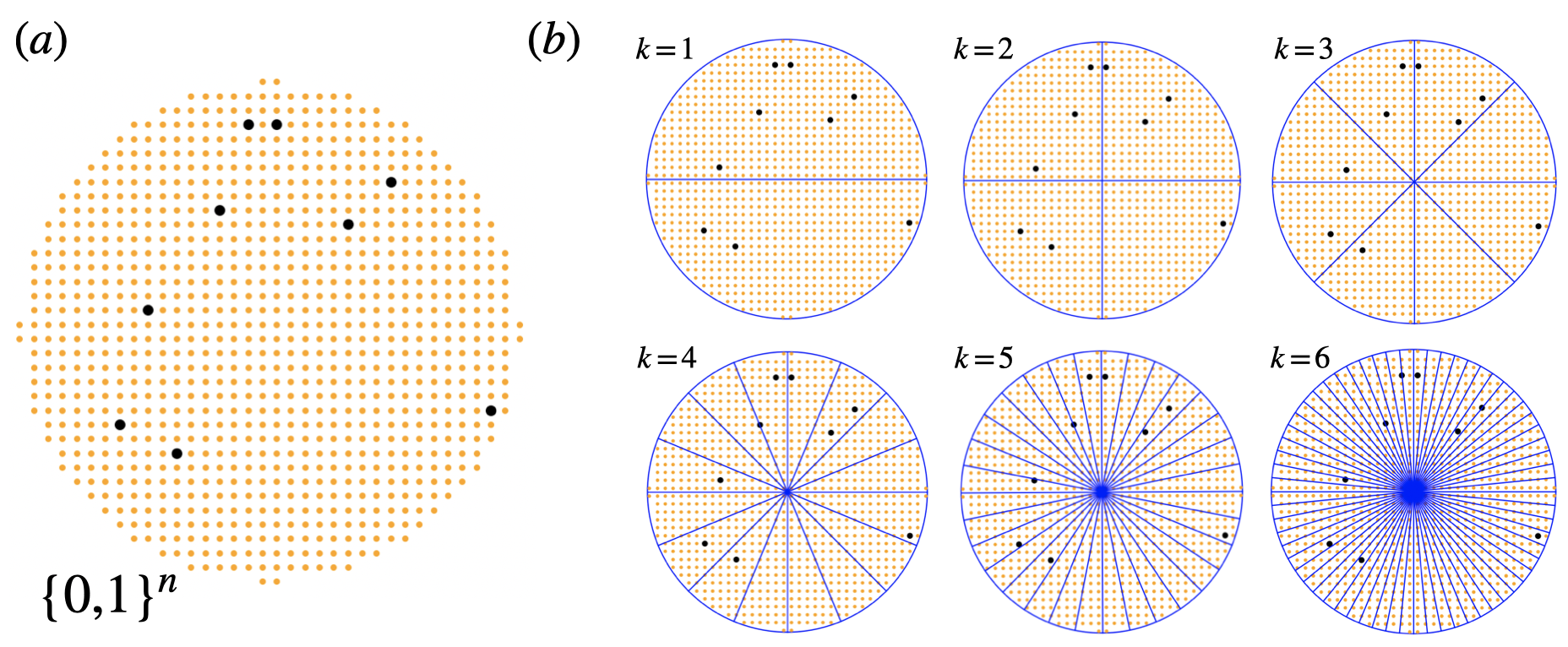} 
\caption{Valiant-Vazirani filter. (a) Illustration of the set $\{0,1\}^n$ containing all $n$-bit strings $x,$ shown as points. Points marked black (chosen randomly)  indicate $s$ satisfying assignments with $f(x)=1$. In this example, $n=10$ and $s=9$. (b) Random partition  into $2^k$ subsets for different values of $k$. For each $k$, the individual sectors can be indexed by a number $ a \in \{ 0,1\}^k$  (not shown in the illustration). When $s \approx 2^k$ there is, on average, 1 satisfying assignment per circular sector.
In the example shown, this occurs when $k = 3$.}
\label{filter figure}
\end{center}
\end{figure} 

\section{\VV reduction}

\subsection{Filtered Boolean formula}

Let $f : \{0,1\}^n \rightarrow \{0,1\}$ be a Boolean function on $n$ bits written in conjunctive normal form (CNF)  with $m$ clauses:
\begin{equation}
f(x) = \bigwedge_{j=1}^m  C_j ,  \ \ C_j = 
\big( l_{j1} \vee  l_{j2}  \vee  l_{j3} \big) .
\label{cnf}
\end{equation}
Here each $l \in \{0,1\}$ is a literal, equal to one of the binary variables  $ x_i$ or its negation $\neg x_i$,  and $m = {\rm poly}(n)$.
We have restricted each clause 
in (\ref{cnf}) to contain exactly three literals;
this is always possible with no loss of generality due to the {\sf NP}-completeness of 3SAT \cite{AroraBarak2009}. Therefore, in what follows, we will  design the \VV algorithm to reduce 3SAT to torsion-based state discrimination. 
Let 
\begin{equation}
S = \bigg\lbrace  x \in \{0,1\}^n : f(x) = 1 
\bigg\rbrace , \ \  s = |S|,
\label{s definition}
\end{equation}
be the set of satisfying assignments, with
$ s \in \{ 0, 1, 2, \cdots , 2^n \}.$ 
The approach of Valiant and Vazirani is to  introduce a filtered Boolean function $\phi(x)$ such that, with constant probability
$p$, its number of satisfying assignments 
$X$  is in the range $ X \in \{ 0,1\},$ 
reducing SAT to UNIQUE SAT \cite{BlassGurevichIC1982}.
The idea behind the filter  is
sketched in Figure~\ref{filter figure}.

For a fixed $k \in \{ 1, \dots ,  n+3\}$, let ${\cal H}$ be the set of pairwise independent hash functions
$ h : \{0,1\}^n \to \{0,1\}^k  $ defined  by
linear equations over $ {\mathbb F}_2 $
(addition  and multiplication are modulo 2):
\begin{equation}
h(x) = Ax + b \ \ {\rm mod}(2) , \ \ 
A \in \{0,1\}^{k \times n} , \ \ b \in \{0,1\}^k.
\label{linear equation for hash}
\end{equation}
 Given $f(x)$, Valiant and Vazirani
construct the filtered function
\begin{equation}
\phi(x) =  f (x) \bigwedge \mathds{1}_{ \{h(x)=0\} } , \ \  \mathds{1}_{ \{h(x)=0\} }  = \big[ h(x) \! = \! 0^k \big] ,
\label{filtered}
\end{equation}
where $h(x)$ is a hash function drawn randomly from ${\cal H}$ with uniform probability, and where $ \big[ A \! = \! B \big]$ is a proposition evaluating to \textsc{True} or \textsc{False}. 

\subsection{Probability of unique solution}

The circuit to evaluate
$\phi(x)$ is described in Section \ref{filtered oracle construction section}.
To explain its operation, we will show that if $s>0$, the random hash constraint will catch exactly one satisfying assignment with probability $p = O(1)$, following the standard pairwise-independent hashing proof of the Valiant--Vazirani reduction~\cite{ValiantVaziraniTCS1986,08104840} However if $f$ is unsatisfiable then $\phi$ is also unsatisfiable. Assuming $s>0$,
let
\begin{equation}
X =  \big| \lbrace  x \in S :h(x) = 0^k
\rbrace  \big|
\label{x definition}
\end{equation}
be a random variable counting the number of satisfying assignments that also satisfy 
 $h(x) = 0^k.$
 This is shown for each $k$ in Figure~\ref{filter figure}b as the number of  satisfying assignments in the first circular sector 
(indexed by  $a=0^k$). 
Then [recall (\ref{filtered})]
\begin{equation}
X = \sum_{x \in S} \mathds{1}_{ \{h(x)=0\} } .
\end{equation}

Next consider an optimal value of  $k$ such that
$ 8s \le 2^k < 16s$. Because $s$ is unknown, the optimal $k$ is also unknown, so the \VV algorithm tries all possible values 
$ k = 0, 1, 2, \dots ,  n+3$
in succession. Then
\begin{equation}
\frac{1}{16} < \lambda \le \frac{1}{8}, \ \ 
{\rm where} \ \ 
\lambda = \frac{s}{2^k}.
\end{equation}
Since $h$ is uniform,
\begin{equation}
\mathbb{E}(\mathds{1}_{ \{h(x)=0\} } ) = 2^{-k} 
\ \ {\rm and} \ \ 
\mathbb{E}(X)  = \lambda.
\end{equation}
In addition, using pairwise independence, 
\begin{equation}
\mathbb{E}( \mathds{1}_{ \{h(x)=0\} }   \mathds{1}_{ \{h(y)=0\} }   )
= \mathbb{E}( \mathds{1}_{ \{h(x)=0\} }  )  \
\mathbb{E}(\mathds{1}_{ \{h(y)=0\} } )
= 2^{-2k}  \ \ {\rm for} \ x \neq y .
\end{equation}
Then we have
\begin{eqnarray}
\mathbb{E}(X^2) &=&
\sum_{x \in S} 
\mathbb{E}( \mathds{1}_{ \{h(x)=0\} } )
+ \sum_{x \neq y} 
\mathbb{E}( \mathds{1}_{ \{h(x)=0\} }   \mathds{1}_{ \{h(y)=0\} }   ) \\
&=& \frac{s}{2^k} + \frac{s(s-1)}{2^{2k}} 
\le \lambda + \lambda^2.
\end{eqnarray}
Using the second-moment inequality,
\[
\Pr(X > 0) \ge \frac{  \mathbb{E}(X)^2}{\mathbb{E}(X^2)}
= \frac{\lambda}{1 + \lambda}.
\]
Since $\lambda > \frac{1}{16}$,  we obtain the following lower bound on $\Pr(X > 0)$:
\[
\Pr(X > 0) > \frac{1}{17}.
\]

Next we obtain an upper bound on $\Pr(X \ge 2)$.
Let 
\begin{eqnarray}
\mathds{1}_{ \{ X\ge 2 \} }  =
\begin{cases}
1, & X \ge 2 , \\
0, & \text{otherwise},
\end{cases}
\end{eqnarray}
and note that 
\begin{eqnarray}
\mathds{1}_{ \{ X\ge 2 \} }   \le   \frac{X(X-1)}{2} ,
\end{eqnarray}
which is established
by checking cases $X=0,1,$ and $\ge 2$ separately.
Then
\begin{equation}
\Pr(X \ge 2) = \mathbb{E}\!\left( \mathds{1}_{ \{ X\ge 2 \} }  \right)   \le \mathbb{E}\!\left( \frac{X(X-1)}{2} \right)
\le \frac{\lambda^2}{2} .
\end{equation}
Since $\lambda \le \frac{1}{8}$, we obtain the following  upper bound on $\Pr(X \ge 2)$:
\[
\Pr(X \ge 2) \le \frac{1}{128}.
\]
Therefore, for the optimal $k$,
the probability $p = \Pr(X \! = \! 1)$ of a unique solution
is
\[
p = \Pr(X > 0) - \Pr(X \ge 2) > \frac{1}{32},
\]
as required.

\section{Filtered oracle construction}
\label{filtered oracle construction section}

\begin{figure}
\begin{center}
\includegraphics[width=10.0cm]{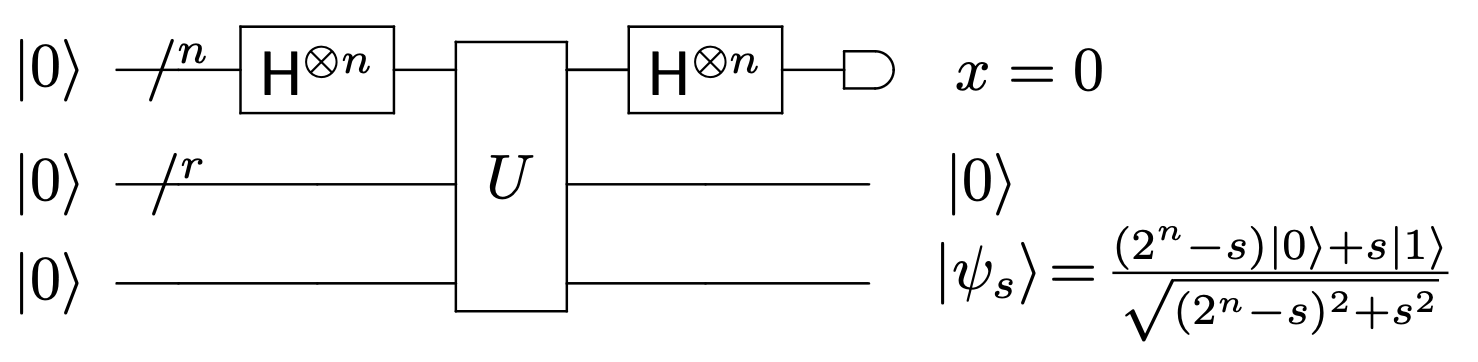} 
\caption{Circuit for encoding the number of solutions $s$ to a Boolean satisfiability problem in an ancilla qubit prior to quantum state discrimination \cite{PhysRevLett.81.3992}.
Here ${\sf H}$ is a Hadamard gate and $U$ implements the Boolean function
$f(x)$ according to (\ref{oracle}).
The postselection step succeeds with probability $\ge \frac{1}{2}$.
}
\label{abrams lloyd figure}
\end{center}
\end{figure} 

Given a Boolean function  $\phi(x)$ on $n$ bits in 3-CNF with $m=O(n^3)$ clauses after removing duplicate clauses, since 3SAT has only $O(n^3)$ distinct three-literal clauses on $n$ variables, an efficient oracle $U$ satisfying
\begin{eqnarray}
U \KET{x} \KET{0}^{ \! \otimes r}  \KET{y}  = \KET{x}  \KET{0}^{ \! \otimes r} \KET{y \oplus \phi(x)}
\label{oracle}
\end{eqnarray}
can be constructed using standard techniques
given in Section \ref{modular circuit construction section}. The first register 
$\KET{x} = \KET{x_1} \KET{x_2}  \cdots \KET{x_n}$ contains $n$ data qubits, the second contains $r$ ancilla qubits initialized to $\KET{0}^{ \! \otimes r}$, used for working space, and the third register is a single qubit holding the output. The encoder is shown in Figure \ref{abrams lloyd figure}.  Applied to a uniform superposition of classical states
$\KET{x}$ and subjected to postselection, the oracle encodes the number of satisfying assignments $s$ into the last qubit for subsequent nonlinear quantum state discrimination \cite{PhysRevLett.81.3992}. 

\subsection{Writing the hash constraint in CNF}

The filtered function (\ref{filtered}) is not itself in CNF, and has to be rewritten in CNF as a preprocessing step. To do this, we can rewrite the indicator function $\mathds{1}_{ \{h(x)=0\} } $
defined in (\ref{filtered}) in CNF.
Using (\ref{linear equation for hash}) we see that
the condition $h(x)=0^k$  is equivalent to a system of linear equations 
\begin{eqnarray}
\bigoplus_{j:A_{ij}=1} \! \!  x_j = b_i, \ \ 
 i=1,\dots ,k, 
\end{eqnarray}
over the field $\mathbb{F}_2$.
Write each row $i$ as
\begin{eqnarray}
x_{j_1}\oplus x_{j_2}\oplus \cdots \oplus x_{j_t}=b_i.
\label{row i}
\end{eqnarray}
For a given row $i$, the $j_1, j_2, \dots , j_t$ are the columns with $A_{ij}=1$.
Next introduce $t-1$ auxiliary variables
 $ z_1,\dots,z_{t-1}$
 to decompose (\ref{row i}) into a sequence of 
 XORs:
\begin{eqnarray}
z_1 = x_{j_1}\oplus x_{j_2}, \ \ 
z_2 = z_1 \oplus x_{j_3},  \ \ 
\cdots  , \ \ 
z_{t-1} = z_{t-2}\oplus x_{j_t}.
\label{unraveling condition}
\end{eqnarray}
In addition there is the overall condition 
\begin{eqnarray}
z_{t-1}=b_i \ \ ({\rm mod} \ 2),
\end{eqnarray}
which can be written in XOR form as
\begin{eqnarray}
z_{t-1} \oplus b_i = 0.
\label{overall condition}
\end{eqnarray}

\begin{figure}
\begin{center}
\includegraphics[width=10.0cm]{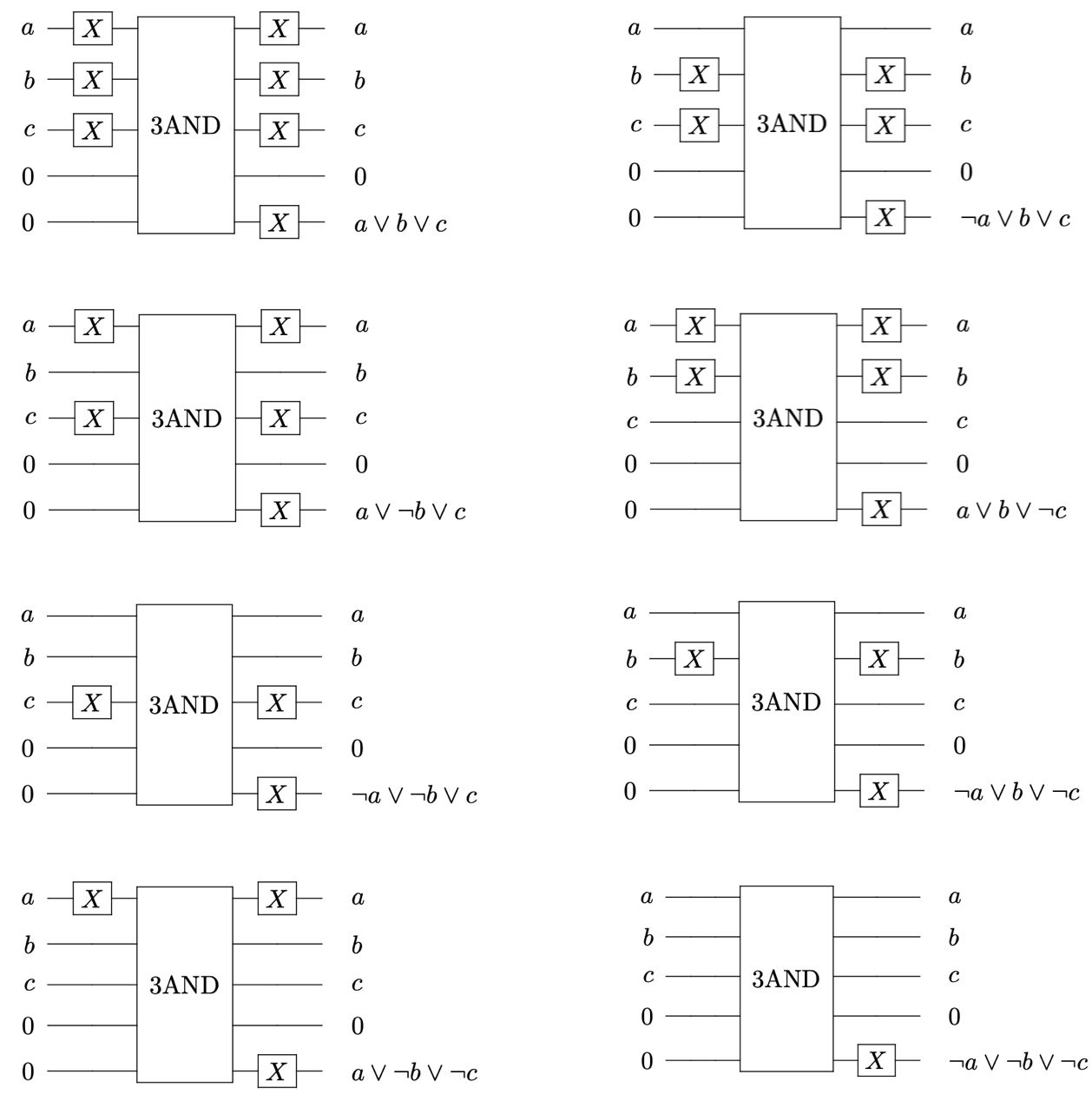} 
\caption{Circuits to implement the 8 clause signatures. Here $a,b,c \in \{0,1\}$ and $X$ is a $\sigma^x$ gate. The 3AND gate is defined in Figure \ref{3and circuit figure}.}
\label{clause signatures figure}
\end{center}
\end{figure} 

For each $i = 1, \dots, k$ there are $t_i$ such XOR propositions (\ref{unraveling condition}) and
(\ref{overall condition}), where $t_i \le n$ is the number of nonzero columns in row $i$ of $A$. 
Each XOR proposition 
\begin{eqnarray}
[a \oplus b = c]
\end{eqnarray}
is equivalent to the CNF formula
\begin{eqnarray}
(\neg a \vee \neg b \vee \neg c)
\;\wedge\;
(a \vee b\vee \neg c)
\;\wedge\;
(a \vee \neg b \vee c)
\;\wedge\;
(\neg a \vee b \vee c)
\end{eqnarray}
with four clauses.
Let $m'$ denote the total number of 3SAT clauses required to implement the hash constraint. In our implementation,
\begin{eqnarray}
m^\prime = 4\sum_{i=1}^k t_i \le 4kn \le 4n(n+3)=O(n^2).
\end{eqnarray}
Finally, form the conjunction of all such clauses with the original formula
\begin{equation}
\phi(x) =  f (x) \bigwedge C^h_{1} \wedge C^h_2 \wedge \cdots \wedge  C^h_{m^\prime} ,
\label{filtered cnf}
\end{equation}
where $C^h$ denotes a clause originating from the hash constraint.

\subsection{Modular circuit construction}
\label{modular circuit construction section}

\begin{figure}
\begin{center}
\includegraphics[width=5.0cm]{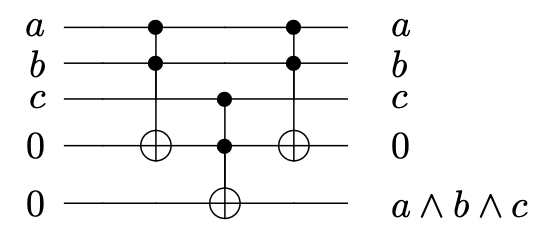} 
\caption{Circuit for the 3AND gate in terms of CCNOT (Toﬀoli) gates.}
\label{3and circuit figure}
\end{center}
\end{figure} 

\begin{figure}
\begin{center}
\includegraphics[width=10.0cm]{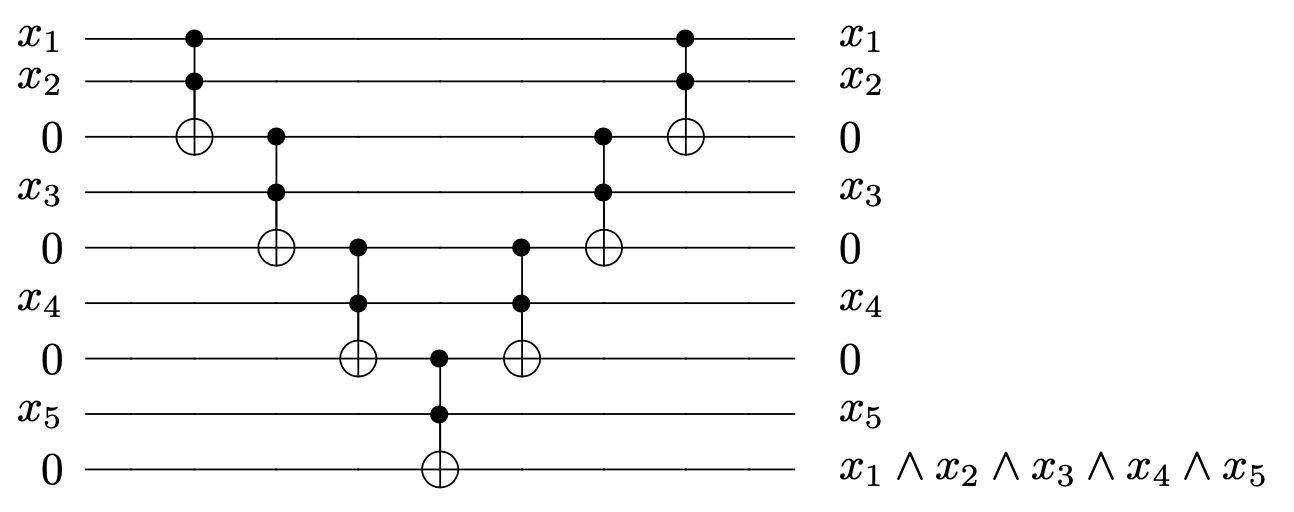} 
\caption{Circuit to implement $M$-bit conjunction 
using  CCNOTs. Here $x_i \in \{0,1\}.$ This requires $O(M)$ CCNOT gates and $O(M)$ additional ancilla initialized to $\KET{0}$. The case $M=5$ is shown.}
\label{ands_circuit figure}
\end{center}
\end{figure}

We have rewritten the filtered function in standard 3SAT form, but now with $M = m + m^\prime$ clauses:
\begin{equation}
\phi(x) = C_{1} \wedge \cdots \wedge  C_{m} 
\wedge C^h_{1} \wedge  \cdots \wedge  C^h_{m^\prime} .
\label{filtered cnf alt}
\end{equation}
Here $m$ is the number of clauses in the original  3SAT formula (\ref{cnf}), and $m^\prime$ clauses are used to implement the hash constraint.
Because every 3SAT clause 
$  l_{1} \vee  l_{2}  \vee  l_{3} $
combines three literals via OR gates, the structure of a clause and its corresponding quantum circuit is determined solely by the placement of negations, which we call the clause signature.
The 8 types are shown in Figure \ref{clause signatures figure}. The circuits are written in terms of the triple AND gate defined in Figure \ref{3and circuit figure}.  An efficient oracle circuit $U$ is then constructed as follows. For each clause $C$:
\begin{enumerate}

 \item[(i)] Copy the required classical inputs $a$, $b$, $c$ to ancilla qubits  in the second register of Figure \ref{abrams lloyd figure}, previously  initialized to $\KET{0}$.
 
 \item[(ii)]  Use two more ancilla 
 (qubits 4 and 5 in Figure \ref{3and circuit figure})
 and implement the circuit from Figure~\ref{clause signatures figure} for the required clause signature.
 
\end{enumerate}
The $M =m + m^\prime$ clauses can be processed in any order or in parallel. The final conjunction of each output (qubit 5 in Figure \ref{clause signatures figure}) is  implemented with a ladder of CCNOTS
as explained in Figure \ref{ands_circuit figure}.
This mirrored CCNOT ladder restores the intermediate ancilla qubits to $\KET{0}$, as required by the oracle definition in (\ref{oracle}).

The resulting circuit to implement the filtered Boolean function for fixed $k$ requires $O(M)$ ancilla and $O(M)$  CCNOT gates.
Trying all $k \in \{1,\dots,n+3\}$ and repeating
$O(1/p)$ times yields a randomized polynomial-time reduction from SAT to UNIQUE-SAT using
$O(n^3)$ ancilla and $O(n^3)$  CCNOT gates per filtered oracle.

\clearpage

\section{Conclusion}

We constructed a quantum implementation of the Valiant--Vazirani reduction for 3SAT. The construction uses a pairwise independent linear hash constraint to isolate a single satisfying assignment with constant probability, while preserving unsatisfiability. Converting the hash constraint to CNF and combining it with the original 3SAT formula gives a filtered 3SAT instance, which we implement as a reversible oracle with polynomial overhead requiring $O(n^3)$ ancilla qubits and $O(n^3)$ CCNOT gates per filtered instance.

When this filtered oracle is combined with torsion-based state discrimination~\cite{150706334,241022032,260514822}, the UNIQUE SAT case can be distinguished in polynomial time in the ideal, noise-free limit. 
Thus, the construction closes the gap between state discrimination using torsion nonlinearity and a polynomial-time solution for {\sf NP} decision problems. The quantum Valiant--Vazirani reduction itself is randomized and polynomial-time, and is not faster than the classical reduction. However, the potential computational advantage is harnessed when coupling a fault-tolerant implementation of the oracle to a nonlinear quantum coprocessor that is also error corrected.

\acknowledgements

This work was partly supported by the NSF under grant no.~DGE-2152159. It is a pleasure to thank Liming Cai for useful discussions.




\bibliographystyle{unsrtnat}
\ifUSEBBL
\bibliography{MS2.bbl}
\else
\bibliography{/Users/mgeller/Dropbox/bibliographies/CM,/Users/mgeller/Dropbox/bibliographies/MATH,/Users/mgeller/Dropbox/bibliographies/QFT,/Users/mgeller/Dropbox/bibliographies/QI,/Users/mgeller/Dropbox/bibliographies/group,/Users/mgeller/Dropbox/bibliographies/books}
\fi

\end{document}